\documentclass[sigconf]{acmart}

\usepackage{listings}
\usepackage{xcolor}
\usepackage{graphics}
\usepackage{lipsum}
\usepackage{subfig}
\usepackage{caption}
\usepackage{wrapfig}
\usepackage{booktabs} 
\usepackage{balance, makecell, geometry}
\usepackage[ruled,norelsize]{algorithm2e}
\makeatletter
\newcommand{\removelatexerror}{\let\@latex@error\@gobble}
\makeatother
\usepackage{amsmath}

\colorlet{punct}{red!60!black}
\definecolor{background}{HTML}{EEEEEE}
\definecolor{delim}{RGB}{20,105,176}
\colorlet{numb}{magenta!60!black}

\lstdefinelanguage{json}{
    basicstyle=\normalfont\ttfamily,
    numberstyle=\scriptsize,
    stepnumber=1,
    numbersep=8pt,
    showstringspaces=false,
    breaklines=true,
    frame=lines,
    backgroundcolor=\color{background},
    literate=
     *{0}{{{\color{numb}0}}}{1}
      {1}{{{\color{numb}1}}}{1}
      {2}{{{\color{numb}2}}}{1}
      {3}{{{\color{numb}3}}}{1}
      {4}{{{\color{numb}4}}}{1}
      {5}{{{\color{numb}5}}}{1}
      {6}{{{\color{numb}6}}}{1}
      {7}{{{\color{numb}7}}}{1}
      {8}{{{\color{numb}8}}}{1}
      {9}{{{\color{numb}9}}}{1}
      {:}{{{\color{punct}{:}}}}{1}
      {,}{{{\color{punct}{,}}}}{1}
      {\{}{{{\color{delim}{\{}}}}{1}
      {\}}{{{\color{delim}{\}}}}}{1}
      {[}{{{\color{delim}{[}}}}{1}
      {]}{{{\color{delim}{]}}}}{1},
}

\copyrightyear{2020}
\acmYear{2020}
\acmConference[WWW '20 Companion]{Companion Proceedings of the Web Conference 2020}{April 20--24, 2020}{Taipei, Taiwan}
\acmPrice{}
\acmDOI{10.1145/3366424.3383534}
\acmISBN{978-1-4503-7024-0/20/04}

\ccsdesc[500]{Information systems~Learning to rank}
\ccsdesc[500]{Information systems~Top-k retrieval in databases}
\ccsdesc[300]{Applied computing~Law, social and behavioral sciences}

\keywords{Ranking, Algorithmic Fairness, Disparate Impact}

\begin{document}

\clubpenalty = 10000
\widowpenalty = 10000

\setlength{\belowdisplayskip}{1pt} 
\setlength{\belowdisplayshortskip}{1pt}
\setlength{\abovedisplayskip}{1pt} 
\setlength{\abovedisplayshortskip}{1pt}

\title{\libname: A Tool For Fairness in Ranked Search Results}

\author{Meike Zehlike}
\affiliation{
	\institution{Humboldt Universit\"at zu Berlin}
	\institution{Max Planck Inst. for Software Systems}
}
\email{meikezehlike@mpi-sws.org}

\author{Tom S\"uhr}
\affiliation{
	\institution{Technische Universit\"at Berlin}
}
\email{tom.suehr@googlemail.com}

\author{Carlos Castillo}
\affiliation{
    \institution{Universitat Pompeu Fabra}
}
\email{chato@acm.org}

\author{Ivan Kitanovski}
\affiliation{%
	\institution{Faculty of Computer Science and Engineering}
	\institution{University Saint Cyril and Methodius}
}
\email{ivan.kitanovski@finki.ukim.mk}

\renewcommand{\shortauthors}{Zehlike et al.}
\newcommand{\libname}{\textsc{FairSearch}\xspace}
\newcommand{\deltr}{\textsc{DELTR}\xspace}
\newcommand{\fair}{\textsc{FA*IR}\xspace}
\newcommand{\spara}[1]{\smallskip\noindent{\bf #1}}
\newcommand{\subsecmargin}{\vspace{-2mm}}
\newcommand{\secmargin}{\vspace{-2mm}}

\begin{abstract}
Ranked search results and recommendations have become the main mechanism by which we find content, products, places, and people online. 
With hiring, selecting, purchasing, and dating being increasingly mediated by algorithms, rankings may determine business opportunities, education, access to benefits, and even social success.
It is therefore of societal and ethical importance to ask whether search results can demote, marginalize, or exclude individuals of unprivileged groups or promote products with undesired features.

In this paper we present \libname, the first fair open source search API to provide fairness notions in ranked search results.
We implement two 
well-known algorithms from the literature, namely \fair~(\citeauthor{zehlike2017fa}, \citeyear{zehlike2017fa}) and \deltr~(\citeauthor{zehlike2018reducing}, \citeyear{zehlike2018reducing}) and provide them as stand-alone libraries in Python and Java. 
Additionally we implement interfaces to Elasticsearch for both algorithms, a well-known search engine API based on Apache Lucene.  
The interfaces use the aforementioned Java libraries and enable search engine developers who wish to ensure fair search results of different styles to easily integrate \deltr and \fair into their existing Elasticsearch environment.
\end{abstract}

\maketitle

\section{Introduction}
\label{sec:introduction}
With the volume of information increasing at a frenetic pace, ranked search results have become the main mechanism by which we find relevant content.
Ranking algorithms automatically score and sort these contents for us, typically by decreasing probability of an item being relevant~\cite{robertson1977probability}.
Therefore, more often than not, algorithms choose not only the products we are offered and the news we read, but also the people we meet, or whether we get a loan or an invitation to a job interview.
With hiring, selecting, purchasing, and dating being increasingly mediated by algorithms, rankings may determine business opportunities, education, access to benefits, and even social success.
It is therefore of societal and ethical importance to ask whether search algorithms produce results that can demote, marginalize, or exclude individuals of unprivileged groups (e.g., racial or gender discrimination) or promote products with undesired features (e.g., gendered books)~\cite{Dwork2012, Sweeney2013, Hardt2014, CaldersChap}.

%
%
This paper operates on the concept of a historically and currently disadvantaged \emph{protected group}, and the concern of \emph{disparate impact}, i.e., a loss of opportunity for said group independently of whether they are treated differently. 
In rankings disparate impact translates into differences in exposure~\cite{singh2018fairness} or inequality of attention across groups, which are to be understood as systematic differences in access to economic or social opportunities.

%
In this paper we present \libname, the first fair open source search API that implements two 
well-known methods from the literature, namely \fair~\cite{zehlike2017fa} and \deltr~\cite{zehlike2018reducing}.
For both algorithms the implementation is provided as a stand-alone Java and Python library, as well as interfaces for Elasticsearch,\footnote{\href{https://www.elastic.co/}{https://www.elastic.co/}} a popular, well-tested search engine, which is used by many big brands such as Amazon, Netflix and Facebook.
Our goal with \libname is to provide various approaches for fair ranking algorithms, with a broad spectrum of justice definitions to satisfy many possible fairness policies in various business situations. 
By providing the algorithms as stand-alone libraries in Python and Java \emph{and} for Elasticsearch we make the on-going research on fair machine learning accessible and ready-to-use for a broad community of professional developers and researchers, particularly those working in the realm of human-centric and socio-technical systems, as well as sharing economy platforms.

\secmargin
\section{Theoretical Background}\label{sec:theory}
This section explains the math behind \fair and \deltr and gives examples for their application domain.
\deltr~\cite{zehlike2018reducing}constitutes a so-called \emph{in-processing} approach, that incorporates a fairness term into its learning objective. 
This way it can learn to ignore the protected feature as well as non-protected ones that serve as proxies, such as ZIP code.
\fair~\cite{zehlike2017fa} belongs to the class of post-processing procedures and re-ranks a given search engine result to meet predefined fairness constraints.

\subsecmargin
\subsection{\deltr: A Learning-To-Rank Approach}\label{subsec:theory-deltr}
\begin{figure}[t]
	\vspace{-5mm}
	\centering
	\subfloat[Case where all non-protected elements appear first in the training set\label{fig:synthethic-nonprotected-first}]{\includegraphics[width=0.48\textwidth]{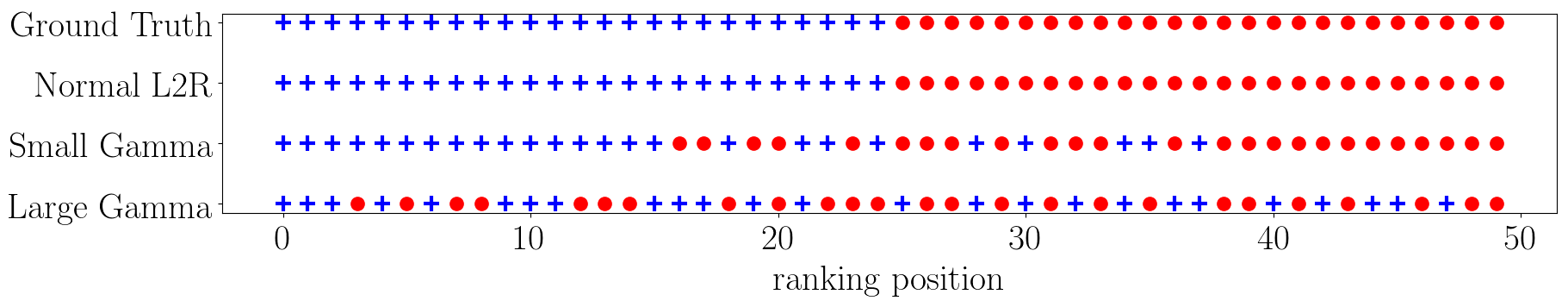}}

	\subfloat[Case where all protected elements appear first in the training set\label{fig:synthetic-protected-first}]{\includegraphics[width=0.48\textwidth]{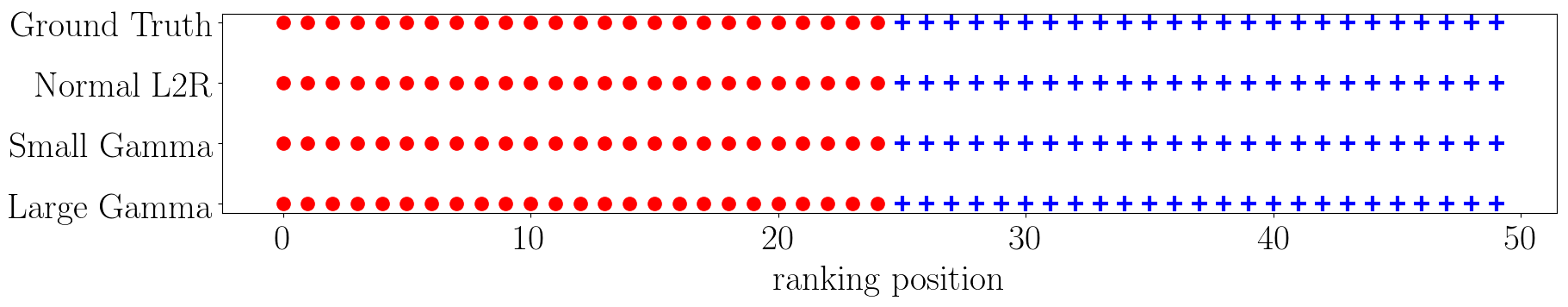}}
	\vspace{-2mm}
	\caption{Depiction of test results using synthetic data.
		Top: \deltr reduces disparate exposure.
		Bottom: asymmetry in \deltr, which does not change rankings if protected elements already appear in the first positions.}
	\vspace{-5mm}
	\label{fig:synthetic-deltr}
\end{figure}

In traditional learning-to-rank (LTR) systems a ranking function $ f $ is learned by minimizing a loss function $ L $, that measures the error between predictions $ \hat{y} $ made by $ f $ and the training judgments $ y $.
For \deltr the loss function of ListNet~\cite{cao2007learning}, a well-known LTR algorithm is extended by a term $ U $, which measures the ``unfairness'' of a predicted ranking. 
This way a new loss function $ L_{\deltr}  = L(y, \hat{y}) + \gamma U(\hat{y}) $ \emph{simultaneously} optimizes $ f $ for relevance \emph{and} fairness.
$ U $ is defined to be a measure of \emph{disparate exposure} across different social groups in a probabilistic ranking $ P_{\hat{y}} $ .
This means discrepancies in the probability to appear at the top position, received by items of the protected group $ G_1 $ vs items of the non-protected group $ G_0 $ are measured: 
\[U(\hat{y}) = \max \left(0, \operatorname{Exposure}(G_0|P_{\hat{y}}) - \operatorname{Exposure}(G_1|P_{\hat{y}})\right)^2
\label{eq:exposure} \] 
Figure~\ref{fig:synthetic-deltr} shows how \deltr works on a synthetic dataset which has a total size of 50 items and each item $ x_i $ is represented by two features: their protection status and a score between 0 and 1: $x_i = ( x_{i, 1}, x_{i, 2} )$.
The attribute $x_{i, 1}$ is 1 if the item belongs to the protected group $G_1$, and 0 otherwise.
The scores $x_{i, 2}$ are distributed uniformly at random over two non-overlapping intervals.
%
Training documents are ordered by decreasing scores, hence the top element is the one that has the highest score.

We first consider a scenario in which all protected elements have strictly smaller scores than all non-protected ones (Figure~\ref{fig:synthethic-nonprotected-first}). 
A standard learning to rank algorithm in this case places all non-protected elements above all protected elements, giving them a larger exposure.
Instead, \deltr with increasing values of $\gamma$ reduces the disparate exposure, while still considering the discrepancy in the score values.
Figure~\ref{fig:synthetic-protected-first} shows the asymmetry of the method: if the protected elements already receive larger predicted exposure than the non-protected by ranker $ f $, \deltr will behave like a standard LTR approach.

\begin{table}[h]
	\small
	\begin{tabular}{r|cccccccccccc}
		\diaghead{some text}%
		{p}{k}&
		1 & 2 & 3 & 4 & 5 & 6 & 7 & 8 & 9 & 10 & 11 & 12 \\ \midrule
		0.1      & 0 & 0 & 0 & 0 & 0 & 0 & 0 & 0 & 0 & 0  &  0 &  0 \\
		0.3      & 0 & 0 & 0 & 0 & 0 & 0 & 1 & 1 & 1 & 1  &  1 &  2 \\
		0.5      & 0 & 0 & 0 & 1 & 1 & 1 & 2 & 2 & 3 & 3  &  3 &  4 \\
		0.7      & 0 & 1 & 1 & 2 & 2 & 3 & 3 & 4 & 5 & 5  &  6 &  6 \\
		\bottomrule
	\end{tabular}
	\vspace{+.5mm}
	\caption{Example values of the minimum number of protected items that must appear in the top $k$ positions to pass the ranked group fairness test with $\alpha=0.1$. We call this an \emph{MTable}. Table from~\cite{zehlike2017fa}}
	\label{tbl:ranked_group_fairness_table}
	\vspace{-10mm}
\end{table}

\subsecmargin
\subsection{\fair: A Re-Ranking Approach}\label{subsec:theory-fair}
Being a post-processing method, \fair~\cite{zehlike2017fa} assumes that a ranking function has already been trained and a ranked search result is available. 
Its \emph{ranked group fairness constraint} guarantees that in a given ranking of length $ k $, the ratio of protected items does not fall far below a given $ p $ at \emph{any ranking position}.
\fair translates this constraint into a statistical significance test, using the binomial cumulative distribution function $ F $ with parameters $ p, k $ and $ \alpha $ and declares a ranking as fairly representing the protected group if, for each $ k $ the following constraint holds:
\[F(\tau_p; k, p) > \alpha,\]
where $ \tau_p $ is the actual number of protected items in the ranking under test.
This constraint can now be used to calculate the minimum number of protected items at each ranking position such that the constraint holds (see table~\ref{tbl:ranked_group_fairness_table} with different examples of $ p $). 
As an example consider the ranking in table~\ref{tbl:xing_intro_example} that corresponds to a job candidate search for an ``economist'' in the XING dataset used in~\cite{zehlike2017fa}.
\begin{table}[h]	
	\vspace{-2mm}
	\centering\small\begin{tabular}{ccccc}\toprule
		Position					  & top 10 & top 10  & top 40 & top 40 \\
		\texttt{1 2 3 4 5 6 7 8 9 10} & male & female & male & female \\
		\midrule
		\texttt{f m m m m m m m m m} & 90\% & 10\% & 73\% & 27\% \\
		\bottomrule
	\end{tabular}
	\vspace{+1mm}
	\caption{Example of non-uniformity of the top-10 vs. the top-40 results for query ``economist'' in XING (Jan 2017). Table from~\cite{zehlike2017fa}
		\label{tbl:xing_intro_example}}
	\vspace{-7mm}
\end{table}
We observe that the proportion of male and female candidates keeps changing throughout the top $ k $ positions, 
which in this case disadvantages women by preferring men at the top-10 positions. 
%
%
Suppose that the required proportion of female candidates is $ p=0.3 $, this translates into having at least one female candidate in the top-10 positions. 
Hence the ranking in table~\ref{tbl:xing_intro_example} will be accepted as fair. 
However, if the required proportion is $ p=0.5 $ this translates into needing at least one female candidate in the top-4, two in the top-7 and three in the top-9 positions.
In this case the ranking will be reordered by \fair to meet the fairness constraints.
Furthermore, our library implements the best possible adjustment of the desired significance level $\alpha$. This is necessary, because the test for a representation like in table \ref{tbl:ranked_group_fairness_table} is a multi-hypothesis test.
\secmargin
\section{\libname: The \deltr Plugin}

\begin{figure*}[t]
	\centering
	\subfloat[Architecture of the \fair Elasticsearch Plugin \label{fig:fair-architecture}]{\includegraphics[page=1,width=.5\textwidth]{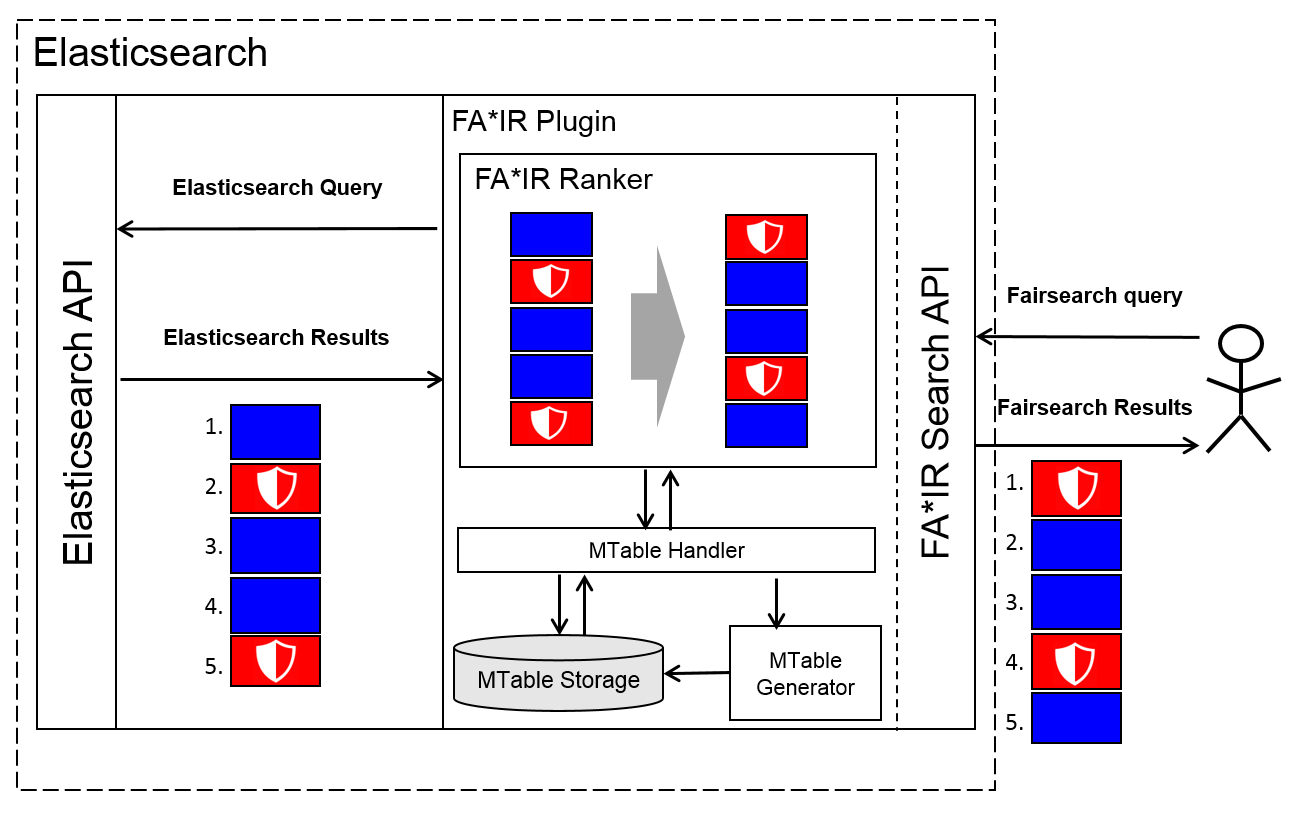}}\hfill
	\subfloat[Demo Application \label{fig:demo_app}]{\includegraphics[page=1,width=.5\textwidth]{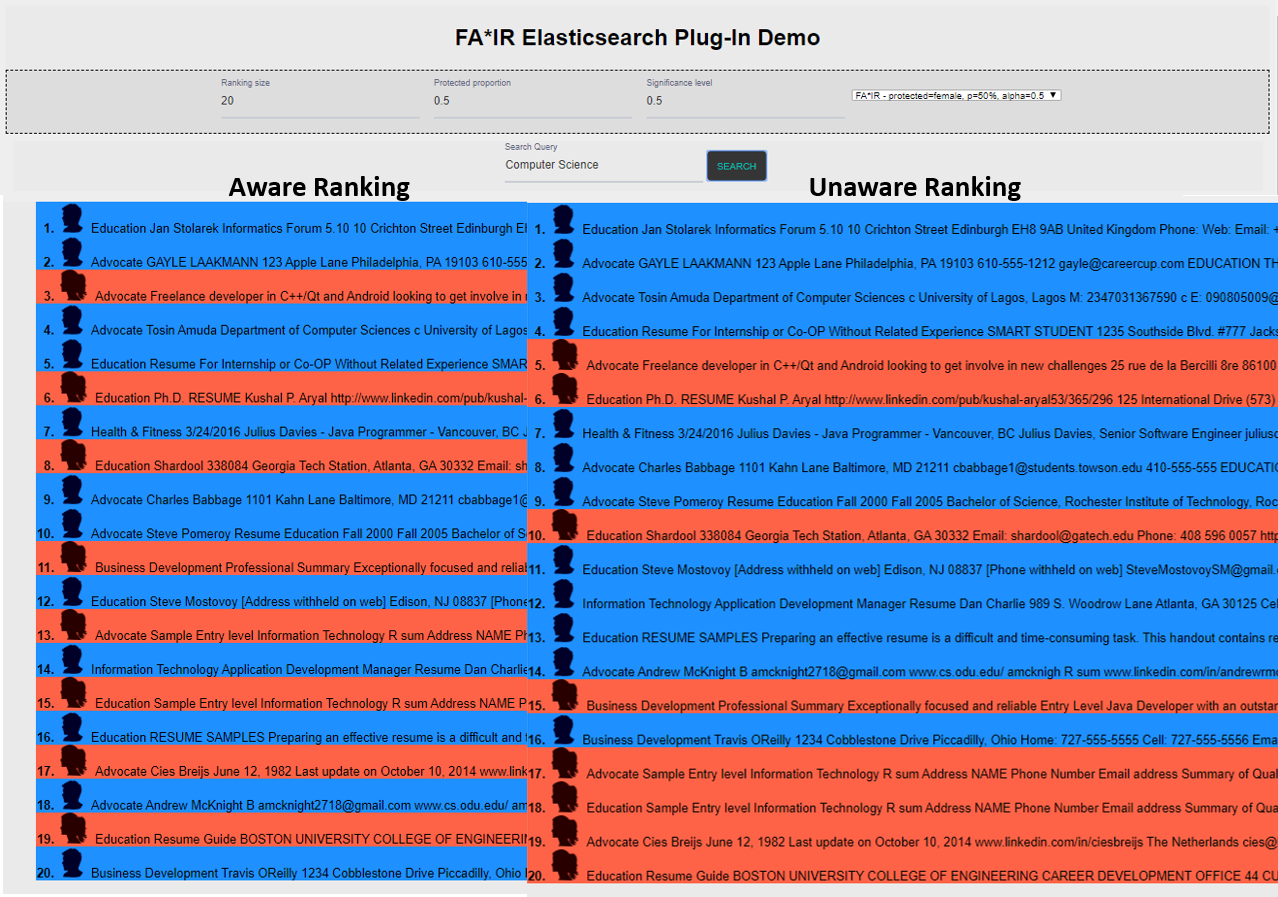}}\hfill
	\vspace{-2mm}
	\caption{ (a) Architecture of the \fair Elasticsearch Plugin and (b) a Demo Webapp with the \fair Elasticsearh Plugin; red indicates protected items} 
	\label{fig:demo_app_and_architecture}
	\vspace{-\baselineskip}
\end{figure*}
\begin{figure}[b]
	\centering
	\includegraphics[width=\columnwidth]{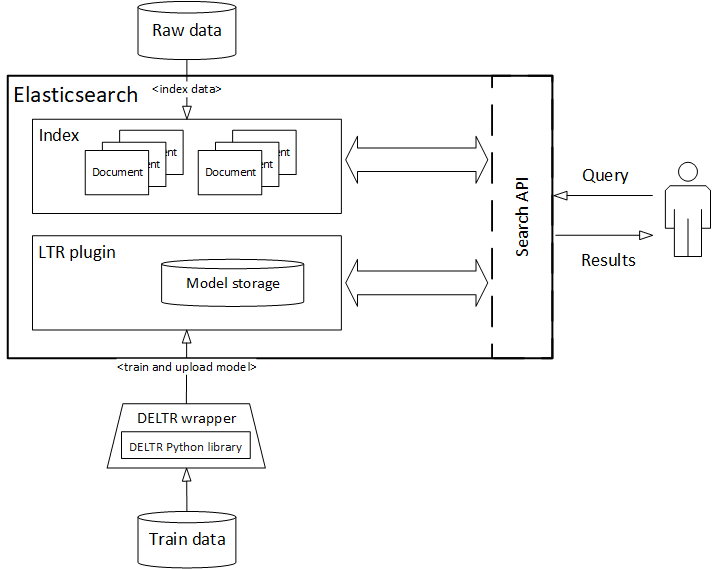}\hfill
	\centering
	\vspace{-3mm}
	\caption{Architecture of the Elasticsearch plugin integration for \deltr} \label{fig:unawareVSFair}
	\vspace{-5mm}
\end{figure}

For the integration of \deltr into Elasticsearch we use the Elasticsearch Learning to Rank (LTR-ES) plugin~\footnote{\url{https://elasticsearch-learning-to-rank.readthedocs.io/en/latest/}}.
%
%
The integration architecture is depicted on Figure~\ref{fig:unawareVSFair}. 
The logic consists of two phases: training and ranking.

\spara{Training.}
To apply \deltr at run-time for retrieval, LTR-ES needs a previously trained model that is uploaded into its model storage.
%
%
Since training models is a very CPU intensive task that involves a lot of supervision and verification, it happens offline in a \deltr wrapper, which calls our stand-alone \deltr Python library to train a LTR-ES suitable model.   
The wrapper has to be provided with a training set, the training parameters and a name for the model. 
%
After training the wrapper calls the LTR-ES upload API, which stores the serialized model inside Elasticsearch's LTR plugin, making it available for up-coming retrieval tasks.
Upon upload the wrapper specifies \texttt{model\_name}, \texttt{type} (always DELTR), the model itself and the \texttt{feature\_set} it was trained against. 
\texttt{feature\_set} specifies query-dependent features, that tell LTR-ES which document features to use when applying the model. 

\spara{Ranking.} 
Elasticsearch ranks retrieved documents by applying re-scoring methods, because executing a query on the entire Elasticsearch cluster is very expensive.
%
%
The system first executes a \textit{baseline relevance} query on the entire index and returns the top $ N $ results. 
The \texttt{Rescorer} then modifies the scores for the top N results and returns the new list. 
\deltr implements Elastic's \texttt{Rescorer} interface, which it applies our previously learned weights to the document features of the top $ N $ results to produce the final ranking.

In the Rescorer, we have to specify two key parameters: 
\begin{itemize}
	\item \texttt{window\_size} - the number of elements to re-score (usually $ N $)
	\item \texttt{model} - the model name.
\end{itemize}
\begin{lstlisting}[language=json,firstnumber=1,basicstyle=\small]
POST someindex/_search
{"query": {
"match": {
"_all": "Jon Snow"}},
"rescore": {
"window_size": 1000,
"query": {
"rescore_query": {
"sltr": {
"params": {
"keywords": "Jon Snow"},
"model": "deltr_model",}}}}}
\end{lstlisting}
The above code constitutes a sample rescore query using \deltr, in which we limit the result set to documents that match ``Jon Snow''. 
All results are scored based on Elasticsearch's default similarity (BM25). 
On top of those already somewhat relevant results we apply our \deltr model to get the best and fairest ranking of the top 1000 documents.

\secmargin
\section{\libname: The \fair Plugin}\label{sec:fairsearch-plugin} 
\begin{figure}[t]
	\vspace{-3mm}
	\includegraphics[width=0.8 \columnwidth]{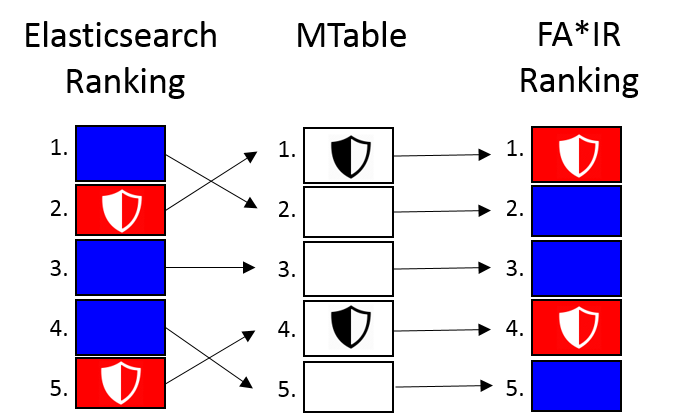}
	\vspace{-2mm}
	\caption{Re-ranking an Elasticsearch result according to a MTable; Shields indicate protected items} \label{fig:fair-reranking}
	\vspace{-4mm}
\end{figure}

The \fair plugin enables Elasticsearch to process a search query and re-rank the result using \fair with parameters $k,p$ and $\alpha$. 
It extends the Elasticsearch API by two new endpoints and a \textit{fair rescorer} JSON object, that contains the parameters for \fair. 
The two new endpoints create a new or request an existing \texttt{MTable}, an integer array that implements table \ref{tbl:ranked_group_fairness_table}. 
Once generated, MTables are persisted within Elasticsearch for further usage to avoid additional computational costs at search time.
Figure \ref{fig:fair-architecture} shows the control flow inside the plugin. 
A \fair query is passed to Elasticsearch, and Elastic returns the standard result ranking to the plugin. 
The plugin then re-ranks the result according to the respective MTable that matches the input parameters $ p, k $ and $ \alpha $. 
Note that the execution of an \textit{unaware} search query with all built-in features is still possible.
\begin{lstlisting}[language=json,firstnumber=1,basicstyle=\small]
POST someindex/_search
{"from" : 0,
"size" : k,
"query" : { "match" : {"body" : q}},
"rescore" : {
"window_size" : k,
"fair_rescorer": {
"protected_key":"gender",
"protected_value":"f",
"significance_level": alpha,
"min_proportion_protected": p}}}
\end{lstlisting}
The components communicate via a REST API for HTTP requests and the above code represents a HTTP request to the plugin.
With this Elasticsearch executes a regular search using the specified \texttt{query} object, the \texttt{match} object and query terms $q$.
%
The result is re-ranked by the plugin using \fair, if the fairness constraints named in $ p, k $ and $ \alpha $ are not met.
First the \texttt{MTable Handler} will check if a MTable for parameters $k,p,\alpha$ already exists (right side of Figure~\ref{fig:fair-architecture}).
If not, the plugin calls the \texttt{MTable Generator} to create it using algorithm~\ref{fig:construct-mtable} and stores it to \texttt{MTable Storage} as key-value pairs with key $(k,p,\alpha)$.
\begin{figure}[t]
	\vspace{-3mm}
	\scalebox{.9}{
		\removelatexerror
		\begin{algorithm}[H]
			\label{fig:construct-mtable}
			\caption{Construct MTable}
			\textbf{INPUT:} Ranking size $k$, minimum proportion $p$, significance $\alpha$;\\
			\textbf{OUTPUT:} MTable $M \in \mathbb{N}^k$\\
			
			$M \leftarrow 0^k$ \;
			$\alpha_c \leftarrow adjustAlpha(k,p,\alpha)$\;			
			\For{$i := 1$ to $k$}
			{
				$M_i \leftarrow inverseCDF(i,p,\alpha_c)$\;
			}
			return $M$\;
		\end{algorithm}  
	}
	\vspace{-7mm}
\end{figure}
We note that the MTable handler in Figure~\ref{fig:fair-architecture} is a simplification of Java classes and interfaces for the purpose of presentation.
The \fair ranker (Figure~\ref{fig:fair-architecture}) re-ranks the Elasticsearch results according to the requested MTable (Figure~\ref{fig:fair-reranking}) and returns them through a HTTP response in JSON format like a standard Elasticsearch result.
\secmargin
\section{Conclusion}

In this paper we presented \libname, the first open source API for search engines to provide fair search results. 
We implemented our previously published methods as stand-alone libraries in Python and Java and embedded those into a plugins for Elasticsearch.
While the plugins are intended to be off-the-shelf implementations for Elasticsearch engineers, the stand-alone libraries allow great flexibility for those who use other technology such as Solr.
%
%
This way we hope that fairness-aware algorithms will make their way faster into productive code and business environments to avoid bad social consequences such as discrimination in search results.

\spara{Acknowledgments.}
This project was realized with a research grant from Data Transparency Lab. Castillo is partially funded by La Caixa project LCF/PR/PR16/11110009. Zehlike is funded by the MPI-SWS. 
\secmargin
\section{Demonstration}
All libraries and plugins are available at \url{https://github.com/fair-search}.
Our demo will consist of two main parts: 
First we will explain the architecture of \fair and \deltr by use of the figures in this paper. 
Next we will have a live coding session.
For \fair we will code a mini example that is going to setup the algorithm in an Elasticsearch instance. 
It will show how to integrate the parameters $p$ and $ \alpha $ and how to further interact with the Elasticsearch plugin via search queries. 
An introduction into the FA*IR python library and Elasticsearch plugin is available on YouTube~\cite{youtubeTutorial}.
For \deltr we will use the synthetic dataset from section~\ref{subsec:theory-deltr} to train a fair model.
We will show how to upload this model into Elasticsearch using the \deltr-Wrapper and how it is used when issuing a search query.

Second using the results from the live coding session we will observe how the algorithms influence ranking results on a demo website (Figure~\ref{fig:demo_app}) for job candidate search, which operates on a resume dataset~\cite{dataset2018}. 
Lastly we will demonstrate how different input parameters for \deltr and \fair will affect the results and give intuition on best practice choices for the parameters.
These two parts are also shown in the YouTube tutorial.

We require a large screen, so that attendees will be able to follow the coding examples from a distance.

\secmargin

{
\bibliographystyle{ACM-Reference-Format}
\bibliography{Main}
}
\balance

\end{document}